\documentclass{article}
\usepackage{spconf,amsmath,graphicx,hyperref}
\usepackage[utf8]{inputenc}
\usepackage[T1]{fontenc}
\usepackage{url}
\usepackage{booktabs}
\usepackage{amsfonts}
\usepackage{nicefrac}
\usepackage{microtype}
\usepackage{xcolor}
\usepackage{multirow}
\usepackage{caption}
\usepackage{enumitem}
\usepackage{etoolbox}

\newcommand{\para}[1]{\noindent\textbf{#1}: }

\makeatletter
\def\bstctlcite#1{%
  \@bsphack
  \@for\@citeb:=#1\do{%
    \edef\@citeb{\expandafter\@firstofone\@citeb\@empty}%
    \if@filesw\immediate\write\@auxout{\string\citation{\@citeb}}\fi
  }%
  \@esphack
}
\makeatother

\AtBeginEnvironment{thebibliography}{\footnotesize}

\makeatletter
\pretocmd{\@makefntext}{\scriptsize}{}{}
\makeatother

\setlist[itemize]{itemsep=2pt, topsep=2pt, parsep=0pt, partopsep=0pt}
\setlist[itemize,1]{leftmargin=1.0em}

\title{Confidence-based Filtering for Speech Dataset Curation with Generative Speech Enhancement Using Discrete Tokens}

\name{Kazuki Yamauchi$^{1,2}$, Masato Murata$^{1}$, Shogo Seki$^{1}$}
\address{$^1$CyberAgent, Japan, $^2$The University of Tokyo, Japan\\
yamauchi-kazuki042@g.ecc.u-tokyo.ac.jp, murata\_masato@cyberagent.co.jp, seki\_shogo@cyberagent.co.jp}

\begin{document}
\ninept
\bstctlcite{BSTcontrol}
\maketitle

\begin{abstract}
Generative speech enhancement (GSE) models show great promise in producing high-quality clean speech from noisy inputs, enabling applications such as curating noisy text-to-speech (TTS) datasets into high-quality ones. However, GSE models are prone to hallucination errors, such as phoneme omissions and speaker inconsistency, which conventional error filtering based on non-intrusive speech quality metrics often fails to detect. To address this issue, we propose a non-intrusive method for filtering hallucination errors from discrete token-based GSE models. Our method leverages the log-probabilities of generated tokens as confidence scores to detect potential errors. Experimental results show that the confidence scores strongly correlate with a suite of intrusive SE metrics, and that our method effectively identifies hallucination errors missed by conventional filtering methods. Furthermore, we demonstrate the practical utility of our method: curating an in-the-wild TTS dataset with our confidence-based filtering improves the performance of subsequently trained TTS models.
\end{abstract}


\section{Introduction}
\label{sec:introduction}

Speech enhancement (SE) is a technology to improve the quality and intelligibility of degraded speech, which is corrupted by background noise, suboptimal recording conditions, or codec artifacts~\cite{zhao2018convolutional}.
Traditional approaches have relied on deep neural networks (DNNs) to estimate complex masks over degraded spectrograms~\cite{xu2014regression,williamson2017time}.
A key application of SE is the cleaning of large-scale, noisy, “in-the-wild” speech datasets, which offer a promising source of diverse speech data~\cite{jung2025titw}.
While in-the-wild datasets are invaluable for developing text-to-speech (TTS) systems, their inherent noise and variability necessitate effective curation~\cite{seki2023text}.
A common strategy for in-the-wild dataset curation~\cite{jung2025titw} follows a two-stage pipeline consisting of: (1) cleaning noisy source speech with an SE model, and (2) subsequently filtering the enhanced speech with a non-intrusive speech quality metric such as DNSMOS~\cite{2021dnsmos}.

Recently, generative SE (GSE) models have emerged~\cite{yang24h_interspeech,koizumi2023miipher,su2021hifi}, showing great promise for producing high-quality clean speech from noisy input by leveraging components developed for TTS systems~\cite{mehta2024matcha,Jungil2020HiFiGAN}.
Building on this progress, we explore the use of GSE to curate in-the-wild datasets into high-quality datasets suitable for TTS training.

Despite this promise, GSE models often introduce characteristic ``hallucination errors,'' such as phoneme omissions and speaker inconsistency~\cite{2023lps}.
These errors can contaminate the training data and ultimately degrade the performance of subsequently trained TTS models, making the filtering of such artifacts a critical step in the curation pipeline.
A key challenge, however, is that conventional filtering methods based on non-intrusive speech quality metrics often fail to detect enhancement-failed speech that, despite sounding acoustically clean, contains GSE-specific hallucination errors~\cite{2023lps}.
While intrusive SE metrics, such as PESQ~\cite{2021pesq}, SpeechBERTScore~\cite{saeki2024speechbertscore}, and LPS~\cite{2023lps}, can detect these errors by comparing the enhanced speech with a clean reference, they are impractical for in-the-wild dataset curation where such references are unavailable.

\begin{figure}[t]
  \begin{center}
  \includegraphics[width=1.0\linewidth]{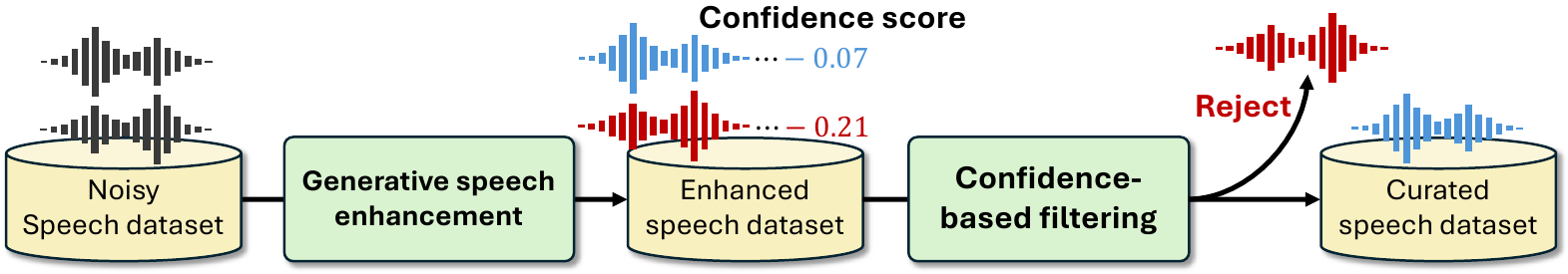}
  \caption{Overview of the speech dataset curation process with our proposed confidence-based filtering.}
\label{fig:curation_overview}
\end{center}
\end{figure}

In this work, we propose a non-intrusive method for filtering hallucination errors from discrete token-based GSE models, as illustrated in Figure~\ref{fig:curation_overview}.
Our method leverages the log-probabilities of generated tokens as confidence scores.
This score serves as a quality metric for filtering potential errors.
Experimental results show that the confidence scores strongly correlate with a suite of intrusive SE metrics.
Furthermore, we show the practical utility of our method by showing that curating an in-the-wild dataset with our confidence-based filtering improves the performance of subsequently trained TTS models.
Our main contributions are summarized as follows:
\begin{itemize}
\item We propose a non-intrusive filtering method that leverages confidence scores derived from discrete token-based GSE models for TTS dataset curation.
\item We show that our filtering method effectively detects GSE-specific hallucination errors missed by conventional filtering methods.
\item We demonstrate that curating a noisy in-the-wild TTS dataset with our confidence-based filtering improves the performance of subsequently trained TTS models.
\end{itemize}

\section{Methods}
\label{sec:methods}

\begin{figure}[t]
  \begin{center}
  \includegraphics[width=1.0\linewidth]{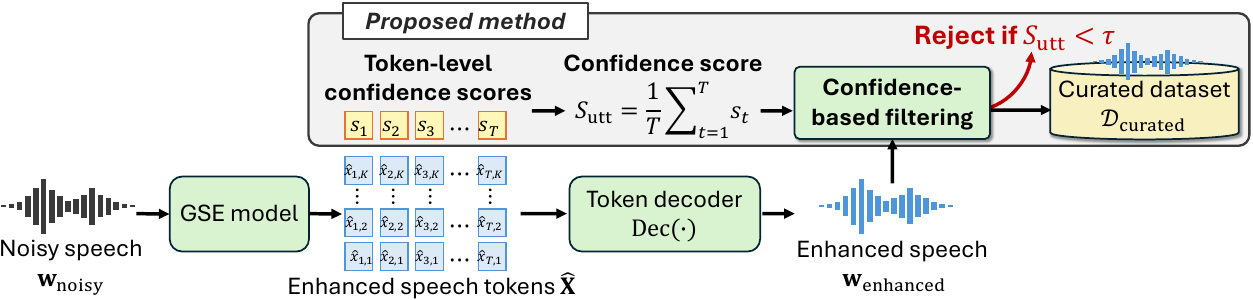}
  \caption{The entire process of our confidence-based filtering. A discrete token-based GSE model outputs enhanced speech $\mathbf{w}_{\text{enhanced}}$, along with token-level confidence scores ($s_1, s_2, \dots, s_T$). If the confidence score $S_{\text{utt}}$ is below a threshold $\tau$, $\mathbf{w}_{\text{enhanced}}$ is filtered out.}
\label{fig:method_overview}
\end{center}
\end{figure}

This section details our method for filtering errors from discrete token-based GSE, as illustrated in Figure~\ref{fig:method_overview}.
Our method leverages confidence scores derived from the probabilities of generated tokens.
These scores are used to detect enhancement-failed speech, enabling effective TTS dataset curation.

\subsection{Discrete audio token-based GSE}
\label{ssec:se_discrete}

A GSE model in a discrete latent space aims to generate a sequence of discrete tokens corresponding to clean speech, conditioned on features extracted from noisy speech.
Specifically, a feature extractor encodes the noisy waveform $\mathbf{w}_{\text{noisy}}$ into conditional features $\mathbf{c}$, and a tokenizer $\text{Tok}(\cdot)$ converts the target clean waveform $\mathbf{w}_{\text{clean}}$ into a sequence of discrete tokens: $\mathbf{X} = \text{Tok}(\mathbf{w}_{\text{clean}})$.
The goal of the GSE model is to generate $\mathbf{X}$ conditioned on $\mathbf{c}$.
To this end, the model is trained to learn the conditional probability distribution $p(\mathbf{X}|\mathbf{c}; \theta)$, where $\theta$ denotes the model parameters.
During inference, a token sequence is sampled from the learned distribution: $\hat{\mathbf{X}} \sim p(\mathbf{X}|\mathbf{c}; \theta)$, which is then fed into a decoder $\text{Dec}(\cdot)$ to synthesize the enhanced speech waveform: $\mathbf{w}_{\text{enhanced}} = \text{Dec}(\hat{\mathbf{X}})$.

\para{Genhancer as a backbone model}
In this work, we employ Genhancer~\cite{yang24h_interspeech} as a backbone SE model.
Genhancer utilizes Descript Audio Codec (DAC)~\cite{kumar2023highfidelity} for both its audio tokenizer $\text{Tok}(\cdot)$ and decoder $\text{Dec}(\cdot)$.
DAC is a neural audio codec based on Residual Vector Quantization (RVQ)~\cite{vasuki2006review}, which quantizes the residual from the previous quantizer, creating a hierarchical representation across $K$ codebooks, each with a vocabulary size of $V$. 
While Genhancer supports several inference strategies, we adopt parallel token prediction, which generates all tokens within a single quantizer in parallel but sequentially across quantizers. Specifically, the conditional distribution over token sequences is factorized as:
\begin{equation}
    p(\mathbf{X}|\mathbf{c}; \theta) = \prod_{k=1}^{K}\prod_{t=1}^{T} p(x_{t, k}|\mathbf{x}_{:, <k}, \mathbf{c}; \theta),
\end{equation}
where $x_{t,k} \in \{1, \dots, V\}$ denotes the token at time step $t$ from the $k$-th quantizer, and $\mathbf{x}_{:, <k}$ represents the token sequences of length $T$ from all preceding quantizers.

\subsection{Confidence scores from discrete token-based GSE}
\label{ssec:estimation}
We propose to leverage log-probabilities of generated tokens as confidence scores to filter out low-confidence outputs that potentially contain errors.
Our method begins by defining a token-level confidence score $s_t$ for each time step $t$.
This score is defined as the log-probability of the generated token $\hat{x}_{t,1}$ from the first quantizer layer ($k=1$), which is chosen as it has the greatest perceptual impact in RVQ-based codecs~\cite{xue2024improving,resynthesiscloser}:
\begin{equation}
    s_t = \log p(x_{t, 1}=\hat{x}_{t, 1} | \mathbf{c}; \theta).
\end{equation}
To obtain a confidence score for the entire utterance, these token-level scores are then averaged over the sequence length $T$ to produce the utterance-level confidence score $S_{\text{utt}}$:
\begin{equation}
    S_{\text{utt}} = \frac{1}{T} \sum_{t=1}^{T} s_t = \frac{1}{T} \sum_{t=1}^{T} \log p(x_{t, 1}=\hat{x}_{t, 1} | \mathbf{c}; \theta).
\end{equation}
We use this final score $S_{\text{utt}}$ as a non-intrusive quality metric to filter the enhanced speech.
A high score indicates that the model generated the output with high confidence, suggesting the enhancement was successful.
Conversely, a low score implies the presence of challenging noise or acoustic distortions that impede the model's processing, flagging the output as a potential enhancement failure.

\subsection{TTS dataset curation with confidence-based filtering}
\label{ssec:curation}
We leverage the utterance-level confidence score $S_{\text{utt}}$ to curate in-the-wild speech datasets into high-quality TTS corpora.
The curation process consists of two stages: (1) cleaning noisy speech with a GSE model and (2) confidence-based filtering.

First, we apply the pre-trained Genhancer model to each utterance in the source speech dataset $\mathcal{D}_{\text{noisy}}$.
For each source speech $\mathbf{w}_{\text{noisy}} \in \mathcal{D}_{\text{noisy}}$, this step yields both an enhanced speech $\mathbf{w}_{\text{enhanced}}$ and its corresponding confidence score $S_{\text{utt}}$.

Second, we perform confidence-based filtering to refine the enhanced dataset $\mathcal{D}_{\text{enhanced}}$.
We define a threshold $\tau$ and retain the enhanced speech $\mathbf{w}_{\text{enhanced}}$ only if its score $S_{\text{utt}}$ exceeds this threshold:
\begin{equation}
    \mathcal{D}_{\text{curated}} = \{ \mathbf{w}_{\text{enhanced}} \in \mathcal{D}_{\text{enhanced}} \mid S_{\text{utt}} \ge \tau \}.
\end{equation}
The threshold $\tau$ can be set empirically, for example, by retaining the top $N\%$ of utterances based on their score distribution.
This filtering step effectively rejects enhancement-failed speech, thereby improving the overall quality of the resultant TTS dataset $\mathcal{D}_{\text{curated}}$.

\begin{table*}[t]
\centering
\caption{SRCC between non-intrusive and intrusive SE metrics on the EARS-WHAM dataset, with ``-in'' metrics calculated on the noisy input speech and ``-out'' metrics on the output speech enhanced by Genhancer. \textbf{Bold} text highlights the best score for each metric.}
\label{tab:correlation}
\scalebox{0.75}{
\begin{tabular}{lccccccc}
\toprule
Metric & ESTOI & SI-SDR & PESQ & SpeechBERTScore & LPS & WAcc & SpkSim \\
\midrule
UTMOS-out & 0.703 & 0.540 & 0.606 & 0.656 & 0.737 & 0.610 & 0.512 \\
UTMOS-in & 0.541 & 0.179 & 0.604 & 0.491 & 0.467 & 0.423 & 0.501 \\
DNSMOS-out & 0.372 & 0.181 & 0.338 & 0.427 & 0.330 & 0.323 & 0.383 \\
DNSMOS-in & 0.673 & 0.381 & 0.720 & 0.614 & 0.569 & 0.546 & 0.639 \\
Whisper confidence-out & 0.728 & 0.529 & 0.676 & 0.736 & 0.770 & \textbf{0.766} & 0.636 \\
Whisper confidence-in & 0.468 & 0.277 & 0.420 & 0.438 & 0.500 & 0.504 & 0.365 \\
CTC score-out & 0.511 & 0.389 & 0.483 & 0.492 & 0.644 & 0.564 & 0.402 \\
CTC score-in & 0.340 & 0.160 & 0.365 & 0.293 & 0.447 & 0.350 & 0.246 \\\midrule
Genhancer confidence (proposed) & \textbf{0.880} & \textbf{0.590} & \textbf{0.883} & \textbf{0.892} & \textbf{0.788} & 0.730 & \textbf{0.790} \\
\bottomrule
\end{tabular}
}
\end{table*}

\section{Related works}
\label{sec:related works}

\para{TTS dataset curation with SE}
Several works have used SE for TTS dataset curation.
For instance, \cite{jung2025titw} constructed a noisy in-the-wild TTS dataset (TITW-hard) and curated it into a cleaner version (TITW-easy). 
Their curation process relied on a discriminative SE model, DEMUCS~\cite{defossez2020real}, and a quality-filtering based on DNSMOS~\cite{2021dnsmos}.
More recently, the LibriTTS-R corpus~\cite{koizumi23_interspeech}, a high-quality TTS corpus, was created by cleaning the LibriTTS corpus~\cite{zen2019libritts} with Miipher~\cite{koizumi2023miipher}, a GSE model.
While this work shows the potential of GSE for TTS dataset curation, LibriTTS is not an in-the-wild dataset.
Consequently, it remains unclear how well GSE performs in more challenging in-the-wild scenarios.
Our work aims to address this gap by investigating the effectiveness of GSE for curating in-the-wild datasets.

\para{In-the-wild dataset screening with quality-filtering}
A different line of work focuses on screening in-the-wild speech datasets without SE.
One approach employs DNN-based quality predictors, such as DNSMOS~\cite{2021dnsmos}, which are trained to directly estimate subjective quality scores from speech samples~\cite{jung2025titw}.
Another approach leverages automatic speech recognition (ASR) systems, for example, by using the connectionist temporal classification (CTC) scores~\cite{kurzinger2020ctc} to filter for intelligible speech~\cite{seki2023text}.
While effective for this initial screening of in-the-wild datasets, the utility of these non-intrusive metrics is limited in a GSE-based curation process, as they are incapable of detecting the hallucination errors introduced by GSE models.

\para{Confidence scores for quality estimation}
The probabilities of generated tokens from discrete token-based generative models are interpreted as a measure of the model's confidence, which is utilized for output quality estimation.
For instance, the perplexity of a language model (LM)—derived from negative log-probabilities—is a standard metric for evaluating generated text quality.
Similarly, confidence scores from ASR models such as Whisper~\cite{radford2022whisper} are used to detect ASR errors~\cite{naderi2024towards}.
For assessing speech generation, SpeechLMScore~\cite{speechlmscore} utilizes the log-probabilities of generated tokens from a pre-trained speech LM.
While our method adopts a similar conceptual approach, it differs in two key aspects.
First, the SE model performs a ``self-assessment'' of its own output, rather than relying on an external speech LM.
Second, our confidence-based metric inherently leverages information from both the input and the output, whereas conventional methods evaluate the output in isolation.

\section{Experiments}
\label{sec:experiments}

We conducted experiments to validate the proposed confidence score as a quality metric for GSE, particularly its ability to detect hallucination errors.
We also evaluated TTS models trained on in-the-wild datasets curated with and without our confidence-based filtering.

\subsection{Experimental setup}
\label{sec:setup}

\subsubsection{Datasets}
\para{GSE training}
We trained Genhancer on public datasets at a 44.1 kHz sampling rate. For training data, we used: 1) clean speech from LibriTTS-R~\cite{koizumi23_interspeech}, upsampled using bandwidth extension~\cite{su2021bandwidth}; 2) noise samples from the TAU Urban Audio-Visual Scenes 2021~\cite{wang2021curated}, DNS Challenge~\cite{dubey2024icassp}, and SFS-Static~\cite{chen2021structure} datasets; and 3) impulse response data from the MIT IR Survey~\cite{traer2016statistics}, EchoThief\footnote{\url{http://www.echothief.com/echothief/}}, and OpenSLR28~\cite{ko2017study}.
Degraded speech for training data was generated using the original configuration from the Genhancer paper.~\cite{yang24h_interspeech}.

\para{Evaluation of filtering methods}
We used EARS-WHAM dataset (v2)\footnote{\url{https://github.com/sp-uhh/ears\_benchmark}} to investigate the correlation between our confidence score and intrusive SE metrics.
This is a simulated dataset created by mixing clean speech from EARS dataset~\cite{richter2024ears} with noise from WHAM! dataset~\cite{wichern2019wham}.
EARS-WHAM consists of approximately 100 hours of simulated noisy speech from 107 speakers and their corresponding clean references that are used to calculate a suite of SE metrics.

\para{Evaluation of subsequently trained TTS models}
We used TITW-hard dataset~\cite{jung2025titw} as our source of in-the-wild TTS data.
This dataset was constructed from the VoxCeleb1 database~\cite{voxceleb}, a large collection of speech data from YouTube.
TITW-hard consists of 282,606 utterances (189 hours) from 1,238 speakers.
We divided them into training (280,130 utterances) and validation (2,476 utterances) sets.

\subsubsection{Models and training}
\para{Discrete token-based GSE model}
We used Genhancer~\cite{yang24h_interspeech} for the GSE model.
We followed the official configuration\footnote{\url{https://sophieymie.github.io/\#training}} for the model architecture and training setup.
We used the pre-trained DAC model~\cite{kumar2023highfidelity}\footnote{\url{https://github.com/descriptinc/descript-audio-codec}} for the audio tokenizer and the layer-wise weighted sum of the pre-trained WavLM model~\cite{chen2022wavlm}\footnote{\url{https://huggingface.co/microsoft/wavlm-large}} to extract conditional features from noisy input.
The model was trained for 400k steps with a batch size of 16 on four NVIDIA A100 GPUs.
During inference, we used a temperature of 0.1 to compute token probabilities.

\para{TTS model}
We used Matcha-TTS~\cite{mehta2024matcha} with the pre-trained HiFi-GAN vocoder (UNIVERSAL\_V1)~\cite{Jungil2020HiFiGAN}\footnote{\url{https://github.com/jik876/hifi-gan}} for the TTS model.
We followed the official implementation of Matcha-TTS\footnote{\url{https://github.com/shivammehta25/Matcha-TTS}} for the model architecture and training setup.
The model was initialized with publicly available weights pre-trained on the VCTK dataset~\cite{vctk}.
Starting from this pre-trained model, we trained it for 500k steps with a batch size of 32 on a single NVIDIA A100 GPU.

\subsubsection{Metrics}
\label{sssec:compared}
\para{Non-intrusive metrics for comparison}
We evaluated the following non-intrusive metrics as data filtering criteria.
The ``-in'' and ``-out'' suffixes denote whether the metric was applied to the noisy input or the enhanced output, respectively.
\begin{itemize}
    \item \textbf{UTMOS-out/in}: A speech naturalness metric that uses the UTMOS model~\cite{saeki22_interspeech} to predict a mean opinion score (MOS).
    \item \textbf{DNSMOS-out/in}: An overall speech quality metric that uses the DNSMOS model~\cite{2021dnsmos} to predict an MOS.
    \item \textbf{Whisper confidence-out/in}: A speech intelligibility metric that uses the mean probabilities of words predicted by a pre-trained Whisper Large v3 model~\cite{radford2022whisper, naderi2024towards}.
    \item \textbf{CTC score-out/in}: A speech intelligibility metric that uses an open-source toolkit\footnote{\url{https://gist.github.com/lumaku/75eca1c86d9467a54888d149dc7b84f1}} to calculate the CTC score~\cite{kurzinger2020ctc, seki2023text}.
    \item \textbf{Genhancer confidence (proposed)}: The confidence score $S_{\text{utt}}$ derived from Genhancer, as described in Section~\ref{ssec:estimation}.
\end{itemize}

\para{Intrusive metrics for evaluation}
We calculated Spearman's Rank Correlation Coefficient (SRCC)~\cite{spearman1987proof} between the non-intrusive metrics described above and the following intrusive SE metrics from the URGENT Challenge 2024~\cite{zhang2024urgent}: ESTOI~\cite{jensen2016algorithm}, SI-SDR~\cite{le2019sdr}, PESQ~\cite{2021pesq}, SpeechBERTScore~\cite{saeki2024speechbertscore}, Levenshtein phoneme similarity (LPS)~\cite{2023lps,zhang2024urgent}, word accuracy (WAcc, equal to ``$1-\text{word error rate (WER)}$''), and speaker similarity (SpkSim).
ESTOI, SI-SDR, and PESQ are traditional intrusive SE metrics for measuring speech signal similarity.
In contrast, SpeechBERTScore and LPS are more recent metrics that evaluate the semantic fidelity of generated speech, with LPS specifically designed to detect hallucination errors in GSE.
To calculate these metrics, we used the official open-source implementation from the URGENT Challenge 2024\footnote{\url{https://github.com/urgent-challenge/urgent2024\_challenge/tree/main/evaluation\_metrics}}.

\subsection{Validation of the confidence score as a GSE metric}
\label{sec:results_correlation}

To validate the proposed confidence score as a quality metric for GSE, we evaluated its correlation with intrusive SE metrics.
We enhanced all samples in the EARS-WHAM dataset using the trained Genhancer model and calculated the metric scores described in Section~\ref{sssec:compared} for each utterance.
Table~\ref{tab:correlation} reports the SRCC between a suite of non-intrusive and intrusive SE metrics.
Among the non-intrusive metrics, the Genhancer confidence score exhibits the strongest correlation with all intrusive metrics except for WAcc.
This shows that the confidence score is a reliable non-intrusive metric for GSE.

\begin{figure}[t]
  \begin{center}
  \includegraphics[width=1.0\linewidth]{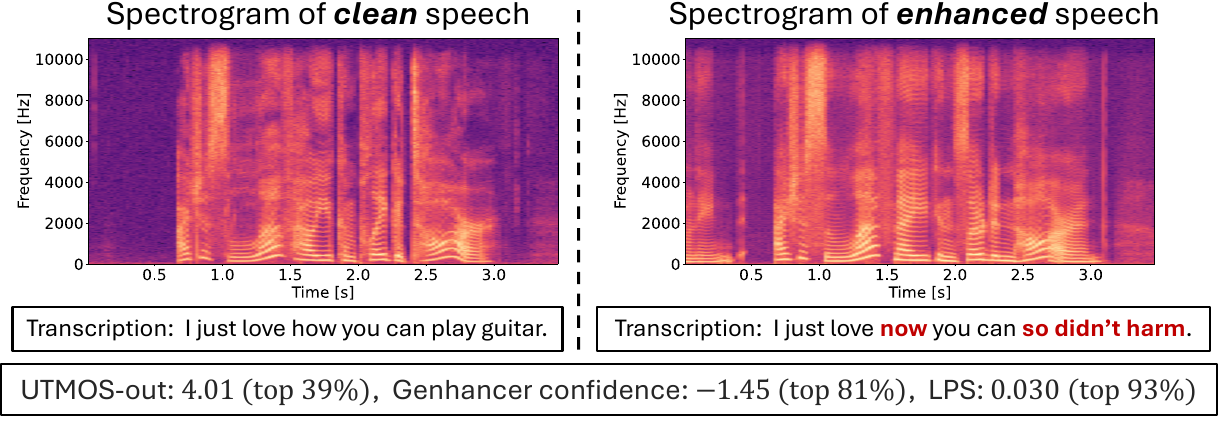}
  \caption{An example of a hallucination error caused by GSE, showing the spectrogram of clean speech (left) and enhanced speech (right).}
\label{fig:spec}
\end{center}
\end{figure}

To demonstrate the ability of our method to detect hallucination errors, Figure~\ref{fig:spec} shows an example of an enhanced speech spectrogram with such errors.
In this example, the enhanced speech sounds acoustically clean but suffers from content corruption, including the generation of meaningless sounds during silent intervals.
In this case, UTMOS-out shows a high score of 4.01 (top $39\%$), suggesting that filtering based on UTMOS-out has difficulty detecting such hallucination errors.
In contrast, both LPS and our confidence score yield relatively low scores, placing them in the top $93\%$ and top $81\%$, respectively.
Note that while LPS is intrusive, making it impractical for in-the-wild dataset curation, our confidence score is non-intrusive and does not require clean reference speech.
This demonstrates that our confidence-based filtering can effectively detect hallucination errors missed by conventional non-intrusive metrics such as UTMOS.

\begin{figure}[t]
  \begin{center}
  \includegraphics[width=1.0\linewidth]{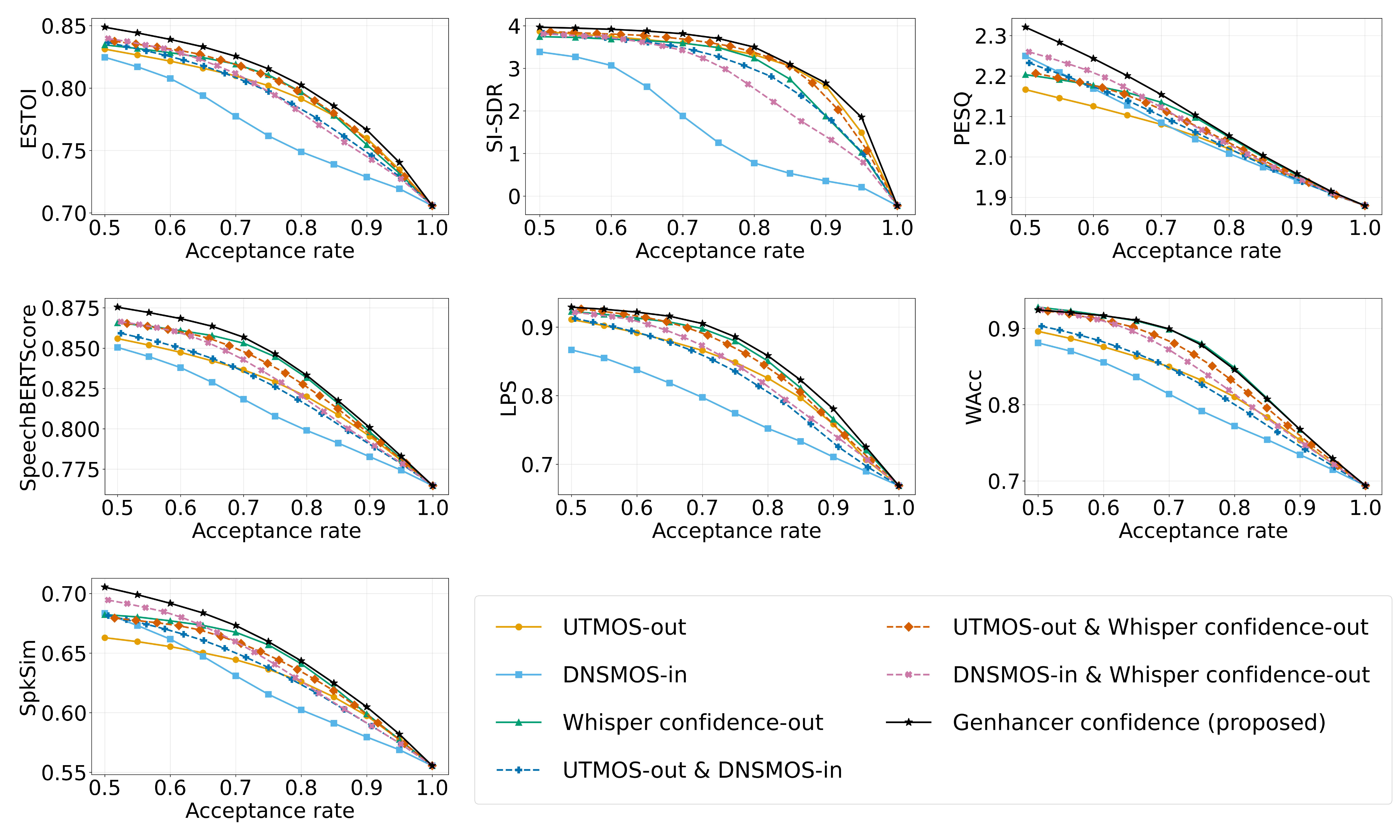}
  \caption{Average quality of the filtered dataset vs. acceptance rate for each filtering method with varying filtering thresholds.}
\label{fig:curves}
\end{center}
          \vspace{-12pt}
\end{figure}

\begin{table}[t]
  \centering
  \caption{Evaluation results of TTS models trained on dataset curated using our proposed method. ``top $N\%$'' indicates the percentage of data with the highest confidence scores that was retained for training. \textbf{Bold} text highlights the best performance for each metric.}
  \label{tab:tts}
  \scalebox{0.82}{
  \begin{tabular}{@{}lcccc@{}}
    \toprule
    & \# Utterances & UTMOS~$\uparrow$ & DNSMOS~$\uparrow$ & WER (\%)~$\downarrow$ \\
    \midrule
    Source (noisy) & 280,130 & 2.73 & 2.74 & 21.31 \\ \midrule
    Enhanced (unfiltered) & 280,130 & 3.64 & 3.10 & 20.45 \\
    Enhanced (top 90\%) & 252,117 & 3.76 & 3.14 & 19.29 \\
    Enhanced (top 80\%) & 224,104 & \textbf{3.80} & \textbf{3.17} & 18.79 \\
    Enhanced (top 70\%) & 196,091 & 3.76 & 3.15 & \textbf{18.14} \\
    Enhanced (top 60\%) & 168,078 & 3.73 & 3.12 & 18.78 \\
    Enhanced (top 50\%) & 140,065 & 3.68 & 3.11 & 20.18 \\
    \bottomrule
  \end{tabular}
  }
\end{table}

\subsection{Analysis of dataset quality vs. quantity}
\label{sec:results_corves}

To examine the trade-off between filtered data quality and quantity, we evaluated filtering methods with varying thresholds on EARS-WHAM.
As baseline filtering criteria, we selected three non-intrusive metrics (UTMOS-out, DNSMOS-in, and Whisper confidence-out) due to their high correlation with intrusive metrics, as shown in Table~\ref{tab:correlation}.
We also evaluated methods combining two metrics, which are denoted by ``\&'' in Figure~\ref{fig:curves}.
For these combined methods, a sample was rejected if its score on either metric fell below the threshold.

Figure~\ref{fig:curves} shows the relationship between the acceptance rate (percentage of retained samples) and the resultant scores on a suite of intrusive SE metrics, averaged over retained samples.
At any given acceptance rate, filtering with Genhancer confidence consistently outperforms the other methods (including combined-metric approaches) across all intrusive metrics except WAcc.
While Whisper confidence-out filtering is competitive with our method on WAcc, its impact on other metrics is marginal.
This indicates that while Whisper confidence primarily focuses on content errors, Genhancer confidence effectively filters for both content and acoustic errors.

\subsection{Evaluation of subsequently trained TTS models}
\label{sec:results_tts}

To demonstrate the practical utility of our method, we evaluated TTS models trained on in-the-wild datasets curated with and without our confidence-based filtering.
First, we enhanced all utterances in TITW-hard using the trained Genhancer model.
Subsequently, we created several subsets by filtering the enhanced dataset based on Genhancer confidence scores. 
This filtering was applied at various thresholds, retaining the top $90\%$, $80\%$, $70\%$, $60\%$, and $50\%$ of the utterances.
We then trained a separate Matcha-TTS model on each of these curated subsets.
For comparison, we also trained two baseline models: one on the noisy source data and another on the enhanced but unfiltered data.
For TTS evaluation, we followed the methodology of the TITW paper~\cite{jung2025titw} except for the text source.
We used each trained model to synthesize speech from 200 randomly selected texts from the SeedTTS \textit{test-en}~\cite{anastassiou2024seed}.
For each of the texts, we randomly selected 40 target speakers from the TITW-hard dataset, generating a total of 8,000 ($=40\times 200$) synthetic utterances per model.
The synthetic speech was then evaluated using UTMOS, DNSMOS, and WER.

Table~\ref{tab:tts} shows the evaluation results for TTS models trained on TITW-hard curated using confidence-based filtering with varying thresholds.
The results show that the model trained on the enhanced dataset curated with our confidence-based filtering outperformed the model trained on the unfiltered enhanced dataset (``Enhanced (unfiltered)'').
Specifically, performance improved as the filtering threshold became stricter, with the best UTMOS and DNSMOS scores (3.80 and 3.17) at the ``top 80\%'' threshold and the lowest WER (18.14\%) at ``top 70\%''. 
However, overly aggressive filtering (e.g., retaining only the top 60\% or 50\%) led to a decline in performance.
These results demonstrate that enhancement errors negatively affect the performance of the subsequently trained TTS model, and confidence-based filtering effectively removes enhancement-failed samples, leading to superior TTS models.
They also suggest a trade-off between data quality and quantity in creating an optimal TTS training dataset.

\section{Conclusion}
\label{sec:conclusion}

In this work, we propose a non-intrusive method for filtering errors from discrete token-based GSE models.
Experimental results show that our method effectively detects GSE-specific hallucination errors, which are frequently overlooked by conventional filtering methods. 
Furthermore, we demonstrate the practical utility of our method: curating an in-the-wild TTS dataset with our confidence-based filtering improves the performance of subsequently trained TTS models.
However, our current approach is limited to discrete token-based GSE models.
To address this issue, future work will explore leveraging the probability distribution of continuous features as a confidence measure for GSE models operating in a continuous latent space~\cite{lu2022conditional,korostik2025modifying}.

\bibliographystyle{IEEEtran}
\bibliography{strings,refs}

@IEEEtranBSTCTL{BSTcontrol,
  CTLuse_forced_etal       = "yes",
  CTLmax_names_forced_etal = "3",
  CTLnames_show_etal       = "3",
}

@INPROCEEDINGS{speechlmscore,
  author={Maiti, Soumi and Peng, Yifan and Saeki, Takaaki and Watanabe, Shinji},
  booktitle={Proc. ICASSP}, 
  title={Speechlmscore: Evaluating Speech Generation Using Speech Language Model}, 
  year={2023},
  pages={1-5},
}

@inproceedings{yang24h_interspeech,
  title     = {Genhancer: High-Fidelity Speech Enhancement via Generative Modeling on Discrete Codec Tokens},
  author    = {Haici Yang and Jiaqi Su and Minje Kim and Zeyu Jin},
  year      = {2024},
  booktitle = {Proc. Interspeech},
  pages     = {1170--1174},
}

@inproceedings{jung2025titw,
  title={The Text-to-speech in the Wild (TITW) Database},
  author={Jung, Jee-weon and Zhang, Wangyou and Maiti, Soumi and Wu, Yihan and Wang, Xin and Kim, Ji-Hoon and Matsunaga, Yuta and Um, Seyun and Tian, Jinchuan and Shim, Hye-jin and others},
  year      = {2025},
  booktitle = {Proc. Interspeech},
  pages     = {4798--4802},
}

@INPROCEEDINGS{2023lps,
  author={Pirklbauer, Jan and Sach, Marvin and Fluyt, Kristoff and Tirry, Wouter and Wardah, Wafaa and Moeller, Sebastian and Fingscheidt, Tim},
  booktitle={Speech Communication; 15th ITG Conference}, 
  title={Evaluation Metrics for Generative Speech Enhancement Methods: Issues and Perspectives}, 
  year={2023},
  pages={265-269},
}

@inproceedings{saeki22_interspeech,
  title={{UTMOS: UTokyo-SaruLab System for VoiceMOS Challenge 2022}},
  author={Takaaki Saeki and Detai Xin and Wataru Nakata and Tomoki Koriyama and Shinnosuke Takamichi and Hiroshi Saruwatari},
  year={2022},
  booktitle={Proc. Interspeech},
  pages={4521--4525},
}

@INPROCEEDINGS{2021dnsmos,
  author={Reddy, Chandan K A and Gopal, Vishak and Cutler, Ross},
  booktitle={Proc. ICASSP}, 
  title={Dnsmos: A Non-Intrusive Perceptual Objective Speech Quality Metric to Evaluate Noise Suppressors}, 
  year={2021},
  pages={6493-6497},
}

@inproceedings{saeki2024speechbertscore,
  author={Saeki, Takaaki and Maiti, Soumi and Takamichi, Shinnosuke and Watanabe, Shinji and Saruwatari, Hiroshi},
  title={Speechbertscore: Reference-aware automatic evaluation of speech generation leveraging nlp evaluation metrics},
  year={2024},
  booktitle={Proc. Interspeech},
  pages={4943--4947},
}

@INPROCEEDINGS{2021pesq,
  author={Rix, A.W. and Beerends, J.G. and Hollier, M.P. and Hekstra, A.P.},
  booktitle={Proc. ICASSP}, 
  title={Perceptual evaluation of speech quality (PESQ)-a new method for speech quality assessment of telephone networks and codecs}, 
  year={2001},
  pages={749-752 vol.2},
}

@inproceedings{radford2022whisper,
    author = {Radford, Alec and Kim, Jong Wook and Xu, Tao and Brockman, Greg and McLeavey, Christine and Sutskever, Ilya},
    title = {Robust Speech Recognition via Large-Scale Weak Supervision},
    booktitle = {Proc. ICML},
    year = {2023},
    pages = {28492--28518}, 
}

@inproceedings{kumar2023highfidelity,
    title={High-Fidelity Audio Compression with Improved {RVQGAN}},
    author={Rithesh Kumar and Prem Seetharaman and Alejandro Luebs and Ishaan Kumar and Kundan Kumar},
    booktitle={Proc. NeurIPS},
    year={2023},
}

@inproceedings{naderi2024towards,
  author={Naderi, Maryam and Hermann, Enno and Nanchen, Alexandre and Hovsepyan, Sevada and Doss, Mathew Magimai},
  title={Towards interfacing large language models with ASR systems using confidence measures and prompting},
  year={2024},
  booktitle={Proc. Interspeech},
  pages={2980--2984},
}

@inproceedings{seki2023text,
  title={Text-to-speech synthesis from dark data with evaluation-in-the-loop data selection},
  author={Seki, Kentaro and Takamichi, Shinnosuke and Saeki, Takaaki and Saruwatari, Hiroshi},
  booktitle={Proc. ICASSP},
  pages={1--5},
  year={2023},
}

@inproceedings{kurzinger2020ctc,
  title={{CTC}-segmentation of large corpora for german end-to-end speech recognition},
  author={K{\"u}rzinger, Ludwig and Winkelbauer, Dominik and Li, Lujun and Watzel, Tobias and Rigoll, Gerhard},
  booktitle={Proc. SPECOM},
  pages={267--278},
  year={2020},
}

@inproceedings{defossez2020real,
  title={Real time speech enhancement in the waveform domain},
  author={Defossez, Alexandre and Synnaeve, Gabriel and Adi, Yossi},
  year={2020},
  booktitle={Proc. Interspeech},
  pages={3291--3295},
}

@inproceedings{koizumi2023miipher,
  title={Miipher: A robust speech restoration model integrating self-supervised speech and text representations},
  author={Koizumi, Yuma and Zen, Heiga and Karita, Shigeki and Ding, Yifan and Yatabe, Kohei and Morioka, Nobuyuki and Zhang, Yu and Han, Wei and Bapna, Ankur and Bacchiani, Michiel},
  booktitle={Proc. WASPAA},
  pages={1--5},
  year={2023},
}

@inproceedings{koizumi23_interspeech,
  title     = {{LibriTTS-R}: A Restored Multi-Speaker Text-to-Speech Corpus},
  author    = {Yuma Koizumi and Heiga Zen and Shigeki Karita and Yifan Ding and Kohei Yatabe and Nobuyuki Morioka and Michiel Bacchiani and Yu Zhang and Wei Han and Ankur Bapna},
  year      = {2023},
  booktitle = {Proc. Interspeech},
  pages     = {5496--5500},
}

@inproceedings{zen2019libritts,
  title={Libritts: A corpus derived from librispeech for text-to-speech},
  author={Zen, Heiga and Dang, Viet and Clark, Rob and Zhang, Yu and Weiss, Ron J and Jia, Ye and Chen, Zhifeng and Wu, Yonghui},
  year={2019},
  booktitle={Proc. Interspeech},
  pages={1526--1530},
}

@article{zhao2018convolutional,
  title={Convolutional neural networks to enhance coded speech},
  author={Zhao, Ziyue and Liu, Huijun and Fingscheidt, Tim},
  journal={IEEE/ACM Transactions on Audio, Speech, and Language Processing},
  volume={27},
  number={4},
  pages={663--678},
  year={2018},
}

@article{xu2014regression,
  title={A regression approach to speech enhancement based on deep neural networks},
  author={Xu, Yong and Du, Jun and Dai, Li-Rong and Lee, Chin-Hui},
  journal={IEEE/ACM transactions on audio, speech, and language processing},
  volume={23},
  number={1},
  pages={7--19},
  year={2014},
}

@article{williamson2017time,
  title={Time-frequency masking in the complex domain for speech dereverberation and denoising},
  author={Williamson, Donald S and Wang, DeLiang},
  journal={IEEE/ACM transactions on audio, speech, and language processing},
  volume={25},
  number={7},
  pages={1492--1501},
  year={2017},
}

@inproceedings{Jungil2020HiFiGAN,
 author = {Kong, Jungil and Kim, Jaehyeon and Bae, Jaekyoung},
 booktitle = {Proc. NeurIPS},
 pages = {17022--17033},
 title = {{HiFi-GAN}: Generative Adversarial Networks for Efficient and High Fidelity Speech Synthesis},
 year = {2020},
}

@inproceedings{su2021hifi,
  title={HiFi-GAN-2: Studio-quality speech enhancement via generative adversarial networks conditioned on acoustic features},
  author={Su, Jiaqi and Jin, Zeyu and Finkelstein, Adam},
  booktitle={Proc. WASPAA},
  pages={166--170},
  year={2021},
}

@inproceedings{mehta2024matcha,
  title={{Matcha-TTS}: A fast TTS architecture with conditional flow matching},
  author={Mehta, Shivam and Tu, Ruibo and Beskow, Jonas and Sz{\'e}kely, {\'E}va and Henter, Gustav Eje},
  booktitle={Proc. ICASSP},
  pages={11341--11345},
  year={2024},
}

@article{spearman1987proof,
  title={The proof and measurement of association between two things},
  author={Spearman, Charles},
  journal={The American journal of psychology},
  volume={100},
  number={3/4},
  pages={441--471},
  year={1987},
}

@inproceedings{zhang2024urgent,
  title={Urgent challenge: Universality, robustness, and generalizability for speech enhancement},
  author={Zhang, Wangyou and Scheibler, Robin and Saijo, Kohei and Cornell, Samuele and Li, Chenda and Ni, Zhaoheng and Kumar, Anurag and Pirklbauer, Jan and Sach, Marvin and Watanabe, Shinji and others},
  year={2024},
  booktitle={Proc. Interspeech},
  pages={4868--4872},
}

@article{jensen2016algorithm,
  title={An algorithm for predicting the intelligibility of speech masked by modulated noise maskers},
  author={Jensen, Jesper and Taal, Cees H},
  journal={IEEE/ACM Transactions on Audio, Speech, and Language Processing},
  volume={24},
  number={11},
  pages={2009--2022},
  year={2016},
}

@inproceedings{le2019sdr,
  title={SDR--half-baked or well done?},
  author={Le Roux, Jonathan and Wisdom, Scott and Erdogan, Hakan and Hershey, John R},
  booktitle={Proc. ICASSP},
  pages={626--630},
  year={2019},
}

@inproceedings{richter2024ears,
  title={{EARS}: An anechoic fullband speech dataset benchmarked for speech enhancement and dereverberation},
  author={Richter, Julius and Wu, Yi-Chiao and Krenn, Steven and Welker, Simon and Lay, Bunlong and Watanabe, Shinji and Richard, Alexander and Gerkmann, Timo},
  year={2024},
  booktitle={Proc. Interspeech},
  pages={4873--4877},
}

@inproceedings{wichern2019wham,
  title={Wham!: Extending speech separation to noisy environments},
  author={Wichern, Gordon and Antognini, Joe and Flynn, Michael and Zhu, Licheng Richard and McQuinn, Emmett and Crow, Dwight and Manilow, Ethan and Roux, Jonathan Le},
  year={2019},
  booktitle={Proc. Interspeech},
  pages={1368--1372},
}

@inproceedings{xue2024improving,
  title={Improving Audio Codec-based Zero-Shot Text-to-Speech Synthesis with Multi-Modal Context and Large Language Model},
  author={Xue, Jinlong and Deng, Yayue and Han, Yicheng and Gao, Yingming and Li, Ya},
  year={2024},
  booktitle={Proc. Interspeech},
  pages={682--686},
}

@inproceedings{resynthesiscloser,
  title={A Closer Look at Neural Codec Resynthesis: Bridging the Gap between Codec and Waveform Generation},
  author={H. Liu, Alexander and Wang, Qirui and Gong, Yuan and Glass, James},
  year={2024},
  booktitle={NeurIPS 2024 Audio Imagination workshop},
}

@article{vasuki2006review,
  title={A review of vector quantization techniques},
  author={Vasuki, A and Vanathi, Ponnusamy Thangapandian},
  journal={IEEE Potentials},
  volume={25},
  number={4},
  pages={39--47},
  year={2006},
}

@inproceedings{vctk,
  title={SUPERSEDED - CSTR VCTK Corpus: English Multi-speaker Corpus for CSTR Voice Cloning Toolkit},
  author={Veaux, Christophe and Yamagishi, Junichi and MacDonald, Kirsten.},
  year={2017},
  booktitle={The Centre for Speech Technology Research (CSTR)},
}

@article{chen2022wavlm,
  title={Wavlm: Large-scale self-supervised pre-training for full stack speech processing},
  author={Chen, Sanyuan and Wang, Chengyi and Chen, Zhengyang and Wu, Yu and Liu, Shujie and Chen, Zhuo and Li, Jinyu and Kanda, Naoyuki and Yoshioka, Takuya and Xiao, Xiong and others},
  journal={IEEE Journal of Selected Topics in Signal Processing},
  volume={16},
  number={6},
  pages={1505--1518},
  year={2022},
}

@inproceedings{su2021bandwidth,
  title={Bandwidth extension is all you need},
  author={Su, Jiaqi and Wang, Yunyun and Finkelstein, Adam and Jin, Zeyu},
  booktitle={Proc. ICASSP},
  pages={696--700},
  year={2021},
}

@inproceedings{wang2021curated,
  title={A curated dataset of urban scenes for audio-visual scene analysis},
  author={Wang, Shanshan and Mesaros, Annamaria and Heittola, Toni and Virtanen, Tuomas},
  booktitle={Proc. ICASSP},
  pages={626--630},
  year={2021},
}

@article{dubey2024icassp,
  title={{ICASSP} 2023 deep noise suppression challenge},
  author={Dubey, Harishchandra and Aazami, Ashkan and Gopal, Vishak and Naderi, Babak and Braun, Sebastian and Cutler, Ross and Ju, Alex and Zohourian, Mehdi and Tang, Min and Golestaneh, Mehrsa and others},
  journal={IEEE Open Journal of Signal Processing},
  volume={5},
  pages={725--737},
  year={2024},
}

@inproceedings{chen2021structure,
    title={Structure from Silence: Learning Scene Structure from Ambient Sound},
    author={Ziyang Chen and Xixi Hu and Andrew Owens},
    booktitle={5th Annual Conference on Robot Learning},
    year={2021},
}

@article{traer2016statistics,
  title={Statistics of natural reverberation enable perceptual separation of sound and space},
  author={Traer, James and McDermott, Josh H},
  journal={Proceedings of the National Academy of Sciences},
  volume={113},
  number={48},
  pages={E7856--E7865},
  year={2016},
}

@inproceedings{ko2017study,
  title={A study on data augmentation of reverberant speech for robust speech recognition},
  author={Ko, Tom and Peddinti, Vijayaditya and Povey, Daniel and Seltzer, Michael L and Khudanpur, Sanjeev},
  booktitle={Proc. ICASSP},
  pages={5220--5224},
  year={2017},
}

@article{anastassiou2024seed,
  title={Seed-tts: A family of high-quality versatile speech generation models},
  author={Anastassiou, Philip and Chen, Jiawei and Chen, Jitong and Chen, Yuanzhe and Chen, Zhuo and Chen, Ziyi and Cong, Jian and Deng, Lelai and Ding, Chuang and Gao, Lu and others},
  journal={arXiv preprint arXiv:2406.02430},
  year={2024}
}

@article{voxceleb,
    author = {Nagrani, Arsha and Chung, Joon Son and Xie, Weidi and Zisserman, Andrew},
    title = {Voxceleb: Large-scale speaker verification in the wild},
    year = {2020},
    volume = {60},
    number = {C},
    journal = {Computer Science and Language},
}

@inproceedings{lu2022conditional,
  title={Conditional diffusion probabilistic model for speech enhancement},
  author={Lu, Yen-Ju and Wang, Zhong-Qiu and Watanabe, Shinji and Richard, Alexander and Yu, Cheng and Tsao, Yu},
  booktitle={Proc. ICASSP},
  pages={7402--7406},
  year={2022},
}

@inproceedings{korostik2025modifying,
  title={Modifying flow matching for generative speech enhancement},
  author={Korostik, Roman and Nasretdinov, Rauf and Juki{\'c}, Ante},
  booktitle={Proc. ICASSP},
  pages={1--5},
  year={2025},
}

\end{document}